\documentclass[pra,aps,twocolumn,superscriptaddress]{revtex4}
\usepackage{graphicx,latexsym}
\usepackage{amsmath}
\newcommand{\sect}[2]{{\par\it #1.---}{#2}}
\newcommand{\beq}{\begin{equation}}
\newcommand{\eeq}{\end{equation}}
\newcommand{\bqn}{\begin{eqnarray}}
\newcommand{\eqn}{\end{eqnarray}}
\newcommand{\bqns}{\begin{eqnarray*}}
\newcommand{\eqns}{\end{eqnarray*}}
\newcommand{\bary}{\begin{array}}
\newcommand{\eary}{\end{array}}
\newcommand{\non}{\nonumber}

\begin{document}
\title{Charged particle motion in a time-dependent flux-driven ring: \\
an exactly solvable model}

\author{Pi-Gang Luan}
\affiliation{Department of Optics and Photonics, National Central
University, Chung-Li 32054, Taiwan}
\author{Chi-Shung Tang}
\affiliation{Research Center for Applied Sciences, Academia Sinica,
Taipei 11529, Taiwan}

\begin{abstract}
We consider a charged particle driven by a time-dependent flux
threading a quantum ring.  The dynamics of the charged particle is
investigated using classical treatment, Fourier expansion technique,
time-evolution method, and Lewis-Riesenfeld approach.  We have shown
that, by properly managing the boundary conditions, a time-dependent
wave function can be obtained using a general non-Hermitian
time-dependent invariant, which is a specific linear combination of
initial angular-momentum and azimuthal-angle operators. It is shown
that the linear invariant eigenfunction can be realized as a
Gaussian-type wave packet with a peak moving along the classical
angular trajectory, while the distribution of the wave packet is
determined by the ratio of the coefficient of the initial angle to
that of the initial canonical angular momentum. From the
topologically nontrivial nature as well as the classical trajectory
and angular momentum, one can determine the dynamical motion of the
wave packet. It should be noted that the peak position is no longer
an expectation value of the angle operator, and hence the Ehrenfest
theorem is not directly applicable in such a topologically
nontrivial system.
\end{abstract}

\maketitle

\section{Introduction}

A charged particle driven by a time-dependent perturbation in a
quantum system is a nontrivial fundamental
issue~\cite{Wees91,FF96,TC99,LC00,Guedes, Bekkar,MG04,LT05}. One can
access the charged particle wave function by placing it in a quantum
ring threaded by a time-dependent magnetic flux.  The vector
potential ${\bf A}(t)$ associated with the time-dependent flux
$\Phi(t)$ times the charge $q$ leading to a phase shift proportional
to the number of flux quanta penetrating the ring, this is known as
Aharonov-Bohm (AB) effect~\cite{AB59,AS87,CP94}.  In adiabatic
cyclic evolution, Berry~\cite{Berry84} first discovered that there
exists a geometric phase.  Later on, Aharonov and Anandan (AA)
removed the adiabatic restriction to explore the geometric phase for
any cyclic evolution~\cite{AA87}.  Time-dependent fields are also
used to deal with field driven Zener tunneling, in which
nonadiabaticity plays a crucial role~\cite{LB85,GT87,MB89}.

In mesoscopic systems, a number of manifestations of the AB effect
have been predicted and
verified~\cite{Butt83,AB91,AB93,AB98,AB-Nature,BK03}.  On the other
hand, Stern demonstrated that the Berry phase affects the particle
motion in the ring similar to the AB effect, and a time-dependent
Berry phase induces a motive force~\cite{Stern92}. It was found
experimentally~\cite{Levy90} that a quantum ring threaded by a
static magnetic field displays persistent currents oscillating in
period of $\Phi_0 = h/q$, the ratio of Planck constant and charge of
a particle.

In the present work, we consider a noninteracting spinless charged
particle moving cyclically in a quantum ring in the presence of a
time-dependent vector potential. Such a particle motion can be
described by the time-dependent Schr{\"o}dinger equation
\begin{equation}
i\hbar\frac{\partial\psi}{\partial
t}=\hat{H}(t)\psi\label{firsteq}\, ,
\end{equation}
where the Hamiltonian $\hat{H}(t)$ is induced by an external
time-dependent vector potential ${\bf A}(t)$, given by
\begin{eqnarray}
\hat{H}(t)&=&\frac{1}{2m}\left[\hat{{\bf P}}-q{\bf A}(t)\right]^2\non\\
&=&\frac{1}{2I}\left[\hat{L}-qRA(t)\right]^2.
\label{HH}\end{eqnarray}
Here $\hat{{\bf P}}= {\bf e}_\theta\hat{P}_\theta$ is the canonical
momentum operator with ${\bf e}_\theta$ being the unit vector along
the azimuthal angle $\theta$; $\hat{L}=\hat{L}_z=(\hat{\bf r}\times
\hat{\bf P})_z$ is the canonical angular momentum operator in the
$z$ direction; $I=mR^2$ is the moment of inertia of the particle;
${\bf A}(t)=A(t){\bf e}_\theta$ is the vector potential; and $R$ is
the radius of the circular ring.  This time-dependent dynamical
problem can be solved by taking into account the Fourier expansion,
time evolution operator, and Lewis-Riesenfeld (LR)
method~\cite{Lewis,LR}.

\section{A Classical treatment}\label{classical}

We first analyze the time-dependent problem in a classical manner.
The time-varying magnetic flux induces an electric field ${\bf
E}=E{\bf e}_\theta$ such that $E=-\partial A/\partial t$. The
charged particle thus obtain a kinematic momentum increment during
the time interval from $0$ to $t$, namely
\begin{equation}
\Delta p_c=\Delta
(mv)=m[v(t)-v(0)]=-q\left[A(t)-A(0)\right],\label{eq1}
\end{equation}
where $p_c=mv$ is the kinematic momentum. It should be noted that
both $p_c$ and $qA$ are not conservative quantities, while from
Eq.~(\ref{eq1}) we see that the canonical momentum $P_c$ is a
constant of motion:
\begin{equation}
P_c(t)=mv(t)+qA(t)=mv(0)+qA(0)=P_c(0). \label{eq2}
\end{equation}

Comparing the two identities in Eq.~(\ref{HH}), we see that the
result of Eq.~(\ref{eq1}) is equivalent to
\begin{equation}
\Delta
l_c=I\left[\omega(t)-\omega(0)\right]=-\frac{q}{2\pi}\left[\Phi(t)-\Phi(0)\right],
\end{equation}
where $l_c=({\bf r}\times {\bf p}_c)_z=I\omega$ indicates the
kinematic angular momentum, $\omega$ is the angular velocity,
$\Phi$ is the magnetic flux threading the ring, and the fact
\begin{equation}
\Phi(t)=2\pi\,RA(t)
\end{equation}
has been used. Also, Eq.(\ref{eq2}) lead us to obtain the following
relations
\begin{equation}
L_c(t)=l_c(t)+\frac{q}{2\pi}\Phi(t)=l_c(0)+\frac{q}{2\pi}\Phi(0)=L_c(0).
\end{equation}
These identities imply that the {\it canonical angular momentum}
$L_c$, defined by $({\bf r}\times {\bf P}_c)_z$, is also a constant
of motion.

Now we define the {\it writhing number} as
\begin{equation}
n_\Phi(t)\equiv\frac{\Phi(t)}{\Phi_0},
\end{equation}
where $\Phi_0= h/q$ is a flux quantum. We also define
\begin{equation}
L_c\equiv n_0\hbar,\;\;\;\;l_c(t)\equiv n_c(t)\hbar,
\end{equation}
then we have
\begin{equation}
n_c(t)=n_0-n_\Phi(t).
\end{equation}
All of these $n$'s are {\it real} numbers.

Now the angular position of the driven particle is given by
\begin{eqnarray}
\theta_c(t)&=&\theta_0+\int^t_0\frac{n_c(\tau)\hbar}{I}d\tau\non\\
&=&\theta_0+\omega_0t-\int^t_0\frac{n_\Phi(\tau)\hbar}{I}d\tau.
\end{eqnarray} Here $\theta_0$ indicates the initial azimuthal
angle; and $\omega_0=\omega(0)=n_0\hbar/I$ stands for the initial
angular velocity. Below we denote the initial kinematic angular
momentum $l_0 \equiv l_c(0)$ for simplicity.

Hereafter we solve the quantum version of the problem, i.e., Eq.
(1), using three different methods. The classical quantities
$l_c(t)$ and $\theta_c(t)$ will also appear in the expressions of
the wave function. Their roles in the quantum problem will be
further explored.

\section{A Fourier expansion method}

The simplest method for solving the time-dependent flux-driven
problem is the Fourier expansion method. The first thing about the
system we discuss is that the wave function satisfies the periodic
boundary condition:
\begin{equation}
\psi(\theta,t)=\psi(\theta+2\pi,t).\label{prdbc}
\end{equation}
The most general form of $\psi$ for the present problem is thus written as
\begin{equation}
\psi(\theta,t)=\sum^{\infty}_{n=-\infty}c_nf_n(t)e^{in\theta},\label{expansion}
\end{equation}
where the $c_n$'s are appropriate coefficients to be determined by
the initial and the boundary conditions.

Substituting Eq.~(\ref{expansion}) into Eq.~(\ref{firsteq}), we can
find the identity
\begin{eqnarray}
&&\sum^{\infty}_{n=-\infty}i\hbar\,c_n \dot{f}_n(t) e^{in\theta}\non\\
&&=\sum^{\infty}_{n=-\infty}c_nf_n(t)\frac{\left(n\hbar-qRA(t)\right)^2}{2I}e^{in\theta}.
\label{feq}
\end{eqnarray}

Solving Eq.(\ref{feq}), after some procedures we obtain
\begin{equation}
f_n(t)=\exp\left\{-\frac{i}{2I\hbar}\int^{t}_{0}\left[n\hbar-qRA(t)\right]^2dt\right\},
\end{equation}
and thus
\begin{equation}
\psi(\theta,t)=\sum^{\infty}_{n=-\infty}c_n\,
\exp\left\{-\frac{i\hbar}{2I}\int^{t}_{0}\left[n-n_\Phi(\tau)\right]^2d\tau+in\theta\right\}.\label{longtime}
\end{equation}

As a simple example, let us choose
\begin{equation}
c_n=N \exp\left[-\sigma^2(n-n_0)^2-i \theta_0(n-n_0)\right],\label{cn}
\end{equation}
where $N$ indicates an appropriate normalization constant; $\sigma$, $\theta_0$ and $n_0$
are real numbers, and $n$ is an integer.

Substituting Eq.(\ref{cn}) into Eq.(\ref{longtime}),
we get
\begin{eqnarray}
&&\psi(\theta,t)=N
\exp\left[-\frac{i}{\hbar}\int^t_0\frac{l_c^2(\tau)}{2I}d\tau+in_0\theta_c(t)\right]
\label{pw}\\
&&\times\sum^{\infty}_{n=-\infty}
\exp\left\{-\sigma^2\left(1+\frac{it}{T}\right)(n-n_0)^2+in\left[\theta-\theta_c(t)\right]
\right\},\non \end{eqnarray} where $T=2I\sigma^2/\hbar$. Applying
the Poisson summation formula
\begin{equation}
\sum^{\infty}_{n=-\infty}f(n)=
\sum^{\infty}_{n=-\infty}\left(\int^{\infty}_{-\infty}f(x)e^{i2\pi n
x}dx\right)
\end{equation}
on the function \begin{eqnarray}
f(x)=\exp{\left[-\sigma^2\left(1+\frac{it}{T}\right)(x-n_0)^2
+i\left(\theta-\theta_c(t)\right)x\right]},\non \\ \end{eqnarray} we
can obtain an alternative expression \begin{eqnarray}
&&\psi(\theta,t) =N\sqrt{\frac{\pi}{\sigma^2(1+\frac{it}{T})}}
\exp\left[-\frac{i}{\hbar}\int^t_0\frac{l_c^2(\tau)}{2I}d\tau\right]\label{pkt}\\
&&\times\sum^{\infty}_{n=-\infty}
\exp\left[-\frac{\left(\theta-\theta_c(t)+2n\pi\right)^2}
{4\sigma^2\left(1+\frac{it}{T}\right)}+in_0(\theta+2n\pi)\right].\non
\end{eqnarray}

We note that Eq. (\ref{longtime}) is the general solution of the
problem, whereas Eq.(\ref{pw}) and Eq.(\ref{pkt}) are two different
expressions for a special solution defined by the $c_n$ coefficients
of Eq. (\ref{cn}).  When $\sigma^2t/T<1$, it should be noted that
Eq.(\ref{pw}) converges slowly while Eq.(\ref{pkt}) converges
quickly.  This means that in the short time limit $t<T/\sigma^2$,
the wave function is better described by a circulating wave packet.
However, for the case of long time limit $t\gg T/\sigma^2$, Eq.
(\ref{pw}) has fast convergency this is because, in this expression,
only the $n\approx n_0$ terms are important. If we further assume
$n_0$ being an integer, then at large $t$ the wave function
approaches to a circulating plane wave that characterized by $n_0$,
namely
\begin{equation}
\psi(\theta,t)\approx N
\exp\left[-\frac{i}{\hbar}\int^t_0\frac{l_c^2(\tau)}{2I}d\tau+in_0\theta\right].\non
\end{equation}
From these findings we conclude that Eq.(\ref{pkt}) describes the
short time behavior and Eq.(\ref{pw}) describes the long time
behavior of the ring system when the wave function is defined by
Eq.(\ref{cn}).

\section{A Time evolution method}

In this section, we shall present how to get the general solution
shown in the previous section in terms of the time evolution
operator $\hat{U}(t)$. The state $|\psi(t)\rangle$ is connected with
the initial state $|\psi(0)\rangle$ through
\begin{equation}
|\psi(t)\rangle=\hat{U}(t)|\psi(0)\rangle.
\end{equation}
and the wave function $\psi(\theta,t)$ is given by
\begin{equation}
\psi(\theta,t)=\langle\theta|\hat{U}(t)|\psi(0)\rangle,
\end{equation}
where $|\theta\rangle$ is the $\theta$-eigenket in the
Schr\"{o}dinger picture that will be explained later.

To begin with, we introduce the canonical commutator
\begin{equation} [\hat{\theta}(0),\hat{L}(0)]=i\hbar. \end{equation}
From this identity, we have
\begin{equation}
[\hat{\theta}(t),\hat{L}(t)]
=\hat{U}^\dagger(t)[\hat{\theta}(0),\hat{L}(0)]\hat{U}(t)=i\hbar.\label{heis}
\end{equation}

Utilizing Eq.(\ref{heis}) we can derive
\begin{equation}
\frac{d\hat{L}(t)}{dt}=\frac{[\hat{L}(t),\hat{H}(t)]}{i\hbar}=0,
\end{equation}
 we thus obtain the identity $\hat{L}(t)=\hat{L}(0).$ Following
 similar procedure it is easy to obtain
\begin{equation}
\frac{d\hat{\theta}(t)}{dt}=\frac{[\hat{\theta}(t),\hat{H}(t)]}{i\hbar}
=\frac{\hat{L}(0)-n_\Phi(t)\hbar}{I},
\end{equation}
which gives us
\begin{equation}
\hat{\theta}(t)=\hat{\theta}(0)+\frac{\hat{L}(0)t}{I}-\int^t_0\frac{n_\Phi(\tau)\hbar
}{I}d\tau.\label{theta}
\end{equation}
Here we see that the canonical angular momentum is a constant of
motion. This is consistent with the classical results discussed in
Sec.~\ref{classical}.

From the above results we have
\begin{equation}
[\hat{H}(t),\hat{H}(t')]=0\label{hh}
\end{equation}
for any two times $t$ and $t'$. Hence the time evolution operator is
simply given by \begin{eqnarray}
\hat{U}(t)&=&\exp\left[-\frac{i}{\hbar}\int^t_0\hat{H}(\tau)d\tau\right]\non\\
&=&\exp\left[-\frac{i\hbar}{2I}\int^t_0
\left(\frac{\hat{L}(0)}{\hbar}-n_\Phi(\tau)\right)^2
d\tau\right].\label{evolution}
\end{eqnarray}

To proceed further, we define $|n\rangle$ as the eigenket of
$\hat{L}(0)$ obeying
\begin{equation}
\hat{L}(0)|n\rangle=n\hbar |n\rangle.
\end{equation}
Then we assume that $|\theta\rangle$ is an eigenket of
$e^{i\hat{\theta}(0)}$ obeying
\begin{equation}
e^{i\hat{\theta}(0)}|\theta\rangle=e^{i\theta}|\theta\rangle.
\end{equation}
The orthogonal conditions of the two eigenkets can thus be expressed
by
\begin{equation}
\langle m|n \rangle =
\frac{1}{2\pi}\int^{2\pi}_0e^{i(m-n)\theta}d\theta=\delta_{mn}
\end{equation}
and
\begin{equation}
\langle\theta|\theta' \rangle =
\frac{1}{2\pi}\sum^{\infty}_{n=-\infty}e^{in(\theta-\theta')}=\delta(\theta-\theta').
\end{equation}
These two orthogonal conditions can be derived from the closure
relations
\begin{equation}
\sum^{\infty}_{n=-\infty}|n\rangle\langle n|=1,\;\;\;\;\;
\int^{2\pi}_0d\theta\, |\theta\rangle\langle\theta|=1
\end{equation}
and taking into account the definition
\begin{equation}
\langle\theta|n\rangle=\frac{1}{\sqrt{2\pi}}e^{in\theta}=\langle n|\theta\rangle^*.
\end{equation}

It should be noted that both $\theta$ and $\theta'$ are defined in
the interval $[0,2\pi)$. In the coordinate representation, the
$e^{in \theta}$ is an eigenfunction of $\hat{L}_{\rm
rep}=-i\hbar\partial/\partial\theta$ with corresponding eigenvalue
$n\hbar$.  This result can be expressed as
\begin{equation}
\langle\theta|\hat{L}(0)|n\rangle=\hat{L}_{\rm
rep}\langle\theta|n\rangle = n\hbar \langle\theta|n\rangle.
\label{leigen1}
\end{equation}
The wave function $\psi$ now can be calculated:
\begin{eqnarray}
&&\psi(\theta,t) \nonumber \\
&=&\sum^{\infty}_{n=-\infty}\langle \theta
|\hat{U}(t)|n\rangle\langle n
|\psi(0)\rangle\non\\
&=&\sum^{\infty}_{n=-\infty}\frac{\langle n
|\psi(0)\rangle}{\sqrt{2\pi}} e^{-\frac{i\hbar}{2I}\int^t_0
\left[n-n_\Phi(\tau)\right]^2 d\tau+in\theta}.
\label{time}
\end{eqnarray}
If we define
\begin{equation}
c_n=\frac{\langle n
|\psi(0)\rangle}{\sqrt{2\pi}},
\end{equation}
then the result of Eq.(\ref{time}) becomes that of Eq.(\ref{longtime}).

Using the time evolution operator $\hat{U}(t)$, we have indeed
found the general solution of Eq.(\ref{longtime}). Based on the
commutativity of the Hamiltonian operator at different times
(Eq.(\ref{hh})), the $\hat{U}(t)$ operator can be constructed
straightforwardly by simple integration.

\section{The Lewis-Riesenfeld method}

In this section, we briefly review the LR method and then apply it
to solve the present problem. We shall show that LR method is not
directly applicable, however, a simple modification concerning about
the boundary condition makes it applicable to solving the problems
with periodic boundary condition.

Traditionally, to utilize the LR method~\cite{LR} solving a
time-dependent system, we have to find an operator $\hat{Q}(t)$ such
that
\begin{equation}
i\hbar\frac{d \hat{Q}}{dt}=i\hbar\frac{\partial \hat{Q}}{\partial
t}+[\hat{Q},\hat{H}]=0, \label{invariant},
\end{equation}
and then find its eigenfunction $\varphi_\lambda(\theta,t)$ satisfying
\begin{equation}
\hat{Q}(t)\,\varphi_{\lambda}(\theta,t)=\lambda\,\varphi_{\lambda}(\theta,t),\label{eigenphi}
\end{equation}
with $\lambda$ being the corresponding eigenvalue.
A wave function $\psi_\lambda (\theta,t)$ satisfying Eq.(\ref{firsteq}) is
then obtained via the relation
\begin{equation} \label{phaseeq}
\psi_\lambda(\theta,t)=e^{i\alpha_{\lambda}(t)}\,\varphi_{\lambda}(\theta,t),
\end{equation}
where $\alpha(t)$ is a function of time only, satisfying
\begin{equation}
\dot{\alpha}_{\lambda}=\varphi^{-1}_{\lambda}
(i{\partial}/{\partial
t}-\hat{H}/\hbar)\varphi_{\lambda}.\label{alpha1}
\end{equation}
A general solution $\psi$ of Eq.(\ref{firsteq}) is then given by
\begin{equation}
\psi(\theta,t)={\sum_\lambda} g(\lambda)\psi_\lambda(\theta,t),
\end{equation}
where $g(\lambda)$ is a weight function for $\lambda$.

To proceed let us assume the time-dependent invariant operator
$\hat{Q}(t)$ takes the linear form~\cite{Guedes, Bekkar}
\begin{equation}
\hat{Q}(t)=a(t)\hat{L}+b(t)\hat{\theta}+c(t)\, ,
\label{iabc}
\end{equation}
in which $a(t)$, $b(t)$, and $c(t)$ are time-dependent $c$-number
functions to be determined.

Substituting Eq.(\ref{iabc}) into
Eq.(\ref{invariant}) and solving these operator equations, we get
\begin{equation}
a(t)=a_0-\frac{b_0t}{I},\;\;\;\;\;
b(t)=b_0,\label{a0b0}
\end{equation}
\begin{eqnarray}
c(t)&=&c_0+b_0\int^t_0\frac{n_\Phi(\tau)\hbar}{I}d\tau,
\label{c0}
\end{eqnarray}
where $a_0$, $b_0$, and $c_0$ are arbitrary complex constants.
Furthermore, substituting Eqs.~(\ref{a0b0}) and (\ref{c0}) into Eq.(\ref{iabc}), we find
\begin{equation} \hat{Q}(t)=a_0\,\hat{L}(0)+b_0\,\hat{\theta}(0)+c_0
=\hat{Q}(0).\end{equation} In other words, the invariant $\hat{Q}$
in the Heisenberg picture is precisely the linear combination of the
initial canonical angular momentum $\hat{L}(0)$ and the initial
azimuthal angle $\hat{\theta}(0)$ with an arbitrary constant $c_0$.
Note that in our system the $\hat{L}$ operator is also an invariant.

It is interesting to ask how the eigenvalue $\lambda$ evolves in
time.   Multiplying the factor $e^{i\alpha(t)}$ on both sides of
Eq.(\ref{eigenphi}), we get
\begin{equation}
\hat{Q}(t)\,\psi_{\lambda}(\theta,t)=\lambda\,\psi_{\lambda}(\theta,t).\label{eigenpsi}
\end{equation}
Partially differentiating the both sides of Eq.(\ref{eigenpsi})
with respect to time and using Eq. (\ref{invariant}), we
find
\begin{equation}
\lambda(t)=\lambda(0),\label{lmbd}
\end{equation}
thus $\lambda$ is a constant.

To find a solution of Eq.(\ref{firsteq}), we have to solve Eq.(\ref{eigenphi}) first.
By solving Eq.(\ref{eigenphi}), we get
\begin{equation}
\varphi_{\lambda}(\theta,t)=\exp\left[\frac{i}{\hbar} \left(\mu(t)\theta
-\frac{1}{2}\nu(t)\theta^2 \right)  \right],\label{phisol}
\end{equation}
where
\begin{equation}
\mu(t)=\frac{\lambda-c(t)}{a(t)},\;\;\;\;\;\nu(t)=\frac{b_0}{a(t)}.
\end{equation}
Substituting Eq.~(\ref{phisol}) into Eq.~(\ref{alpha1}), we obtain
\begin{equation}
\alpha_{\lambda}(t)=\alpha_\lambda(0)-\int^t_0
\frac{\left[\eta^2(\tau)+i\hbar\nu(\tau)\right]}{2I\hbar} d\tau,\label{mueta}
\end{equation}
where
\begin{equation}
\eta(\tau)\equiv\mu(\tau)-n_\Phi(\tau)\hbar.
\end{equation}
In the derivation of Eq.(\ref{mueta}), we have used the following
two identities:
\begin{equation}
\dot{\mu}=\frac{\nu\left(\mu-n_\Phi\hbar\right)}{I},\;\;\;
\dot{\nu}=\frac{\nu^2}{I}.\label{munu}
\end{equation}
Here we see that in general $\alpha_\lambda(t)$ is a complex function.

Although the form of
$\psi_\lambda(\theta,t)=e^{i\alpha_\lambda(t)}\phi_{\lambda}(\theta,t)$
is indeed a solution of Eq.(\ref{firsteq}), however, it does not
satisfy the periodic boundary condition [see Eq.(\ref{prdbc})]. This
problem can be resolved by defining the total wave function
$\psi(\theta,t)$ as the summation of all
$\psi_\lambda(\theta+2n\pi,t)$ terms: \begin{eqnarray}
&&\psi(\theta,t)=\sum^{\infty}_{n=-\infty}\psi_\lambda(\theta+2n\pi,t)\non\\
&&=\sum^{\infty}_{n=-\infty}
\exp\left[i\alpha_\lambda(t)+\frac{i}{\hbar} \mu(t)(\theta+2n\pi)\right.\non\\
&&\hspace{3cm}\left.-\frac{i}{2\hbar}\nu(t)(\theta+2n\pi)^2 \right].\label{result1}
\end{eqnarray}

It can also be transformed to the equivalent form below using the
Poisson summation formula: \begin{eqnarray} &&\psi(\theta,t)\non
\\
&&=\sqrt{\frac{\hbar}{2\pi i\nu(t)}}\exp\left[i\alpha_\lambda(t)
+i\frac{\nu(t)}{2\hbar}
\theta^2_c(t)+in_0\theta_c(t)\right]\non\\
&&\times\sum^{\infty}_{n=-\infty}
\exp\left[\frac{i\hbar(n-n_0)^2}{2\nu(t)}+in\left(\theta-\theta_c(t)\right)\right].
\label{result2} \end{eqnarray}
From these derivations it should be noted that {\it when using the
LR method, the boundary conditions have to be carefully managed,
otherwise one may get an incorrect result}.

\section{A Comparison to Various Approaches}

In this section we shall show that Eq.~(\ref{result1}) and
(\ref{result2}) can be cast into the forms of Eq.(\ref{pkt}) and
Eq.~(\ref{pw}), respectively. To proceed further, let us first
borrow the parameters $n_0$ and $\theta_c(t)$ from Sec.~II in
combination with the results obtained in Sec.~V, we have the simple
identity
\begin{equation}
a(t)n_0\hbar+b(t)\theta_c(t)+c(t)=a_0n_0\hbar+b_0\theta_0+c_0.
\end{equation}
Comparing this result with Eq.~(\ref{lmbd}), we find that they are
very similar. For simplicity, we define
\begin{equation}
\lambda\equiv a(t)n_0\hbar+b(t)\theta_c(t)+c(t),\label{lmbd1}
\end{equation}

in combination with Eq.(\ref{lmbd1}), it is easy to obtain
\begin{eqnarray}
\mu(t)&=&n_0\hbar+\nu(t)\theta_c(t),\\
\non\\
\eta(t)&=&l_c(t)+\nu(t)\theta_c(t). \end{eqnarray} Further, using
the identity
\begin{eqnarray} \frac{d\theta^2_c(t)}{dt}=\frac{2}{I}l_c(t)\theta_c(t) \end{eqnarray}
 and the identity of $\dot{\nu}$ in Eq.~(\ref{munu}), we have
\begin{equation}
\eta^2=l^2_c+I\frac{d}{dt}(\nu\theta^2_c).\label{eta2}
\end{equation}

Substituting Eq.(\ref{eta2}) into Eq.(\ref{mueta}), we get
\begin{equation}
e^{i\alpha_\lambda(t)}=\frac{e^{i\alpha_\lambda(0)}\exp
\left(-\frac{i}{\hbar}\int^t_0\frac{l^2_c(\tau)}{2I}d\tau-\frac{i\nu\theta^2_c}{2\hbar}\right)
}{\sqrt{1-\frac{\nu_0t}{I}}}.
\end{equation}
In addition, by defining $e^{i\alpha_\lambda(0)}$ and $\nu_0$ as
\begin{equation}
e^{i\alpha_\lambda(0)}\equiv \frac{N\sqrt{\pi}}{\sigma}
\end{equation}
and
\begin{equation}
\nu_0\equiv-\frac{iI}{T}=-\frac{i\hbar}{2\sigma^2},
\end{equation}
we can see clearly that Eqs.~(\ref{result1}) and (\ref{result2})
become exactly the same as Eqs.~(\ref{pkt}) and (\ref{pw}). Hence,
we have verified that the Fourier transform method, the time
evolution method, and the Lewis-Riesenfeld method are equivalent
when we choose the coefficients $c_n$ as Eq.(\ref{cn}).   This
restriction is not necessary to find the two equivalent general
solutions Eq.(\ref{longtime}) and Eq.(\ref{time}) obtained by using
Fourier expansion and time evolution methods, respectively.  It
turns out that the Lewis-Riesenfeld method seemed to be more
restrictive.

It is now interesting to discuss the physical meanings of $l_c(t)$
and $\theta_c(t)$ we have obtained.  Although they are originated
from the classical treatment, however, in what sense do they play a
role of dynamic variables in the corresponding classical system
should be further clarified.  We would like to bring attention that
in the Schr\"{o}dinger picture within coordinate representation, the
$l_c(t)$ is the expectation value of
$\hat{L}-qRA(t)=-i\hbar\partial/\partial\theta-qRA(t)$. However,
$\theta_c(t)$ is not the expectation value of $\hat{\theta}=\theta$
with respect to the wave function obtained in Eq.(\ref{pw}),
instead, it is merely the peak position of the wave packet [see
Eq.(\ref{pkt})].

In other words, the conventional Ehrenfest Theorem is not directly
applicable in this topologically nontrivial system. This
consequence is due to the fact that we are not able to distinguish
the phase between the angle $\theta$ and $\theta+2n\pi$. Hence the
$\hat{\theta}$ operator is not well-defined, only the
$e^{i\hat{\theta}}$ is a well-defined operator, as has been
demonstrated in Sec.~IV. These facts cause the $\lambda$ losing
its meaning as an expectation value of the $\hat{Q}$ operator.

We finally point out the relationship between the problem we have
considered here and that we studied in the previous work
\cite{LT05}. In that work we studied the motion of a charged
particle in a one-dimensional space subject to a time-varying linear
potential. By doing a gauge transformation as that mentioned in
Ref.~\onlinecite{JB}, the Hamiltonian in Ref.~\onlinecite{LT05} can
be cast into the form of Eq.(\ref{firsteq}), hence the two problems
are equivalent if we ignore the difference of their topologies.
Therefore, by ignoring the factor caused by the gauge transformation
(which contains only a function of time), the circulating wave
packet solution, Eq.(\ref{pkt}), can be viewed as the wave packet
solution in a one-dimensional system (see Eqs.(24) and (40) in
Ref.~\onlinecite{LT05}) being folded into a ring. That is why in
Eq.(\ref{result1}) the total wave function $\psi(\theta,t)$ can be
written as the sum of all the $\psi_\lambda(\theta+2n\pi,t)$ terms.
This folding nature of wave packet leads to interferences between
different $\psi_\lambda(\theta+2n\pi,t)$ terms. As a result, the
expectation value of $\hat{\theta}$ operator is different from the
peak position $\theta_c$ of the wave packet.

\section{Concluding Remarks}

In this article, we have studied the problem of a charged particle
moving in a ring subject to a time-dependent flux threading it.
After analyzing the problem in a classical manner, various
approaches including Fourier expansion method, time-evolution
method, and Lewis-Riesenfeld method are considered and compared. In
the Lewis-Riesenfeld approach, by appropriately managing the
periodic boundary condition of the system, a time-dependent wave
function can be obtained by using a non-Hermitian time-dependent
linear invariant. The eigenfunction of the invariant can be realized
as a Gaussian-type wave packet with the peak moving along the
classical angular trajectory, while the distribution of the wave
packet is determined by the ratio of the coefficient of the initial
angle to that of the initial canonical angular momentum. In this
circular system, we find that although the classical trajectory and
angular momentum can determine the motion of the wave packet;
however, the peak position is no longer an expectation value of the
angle operator, and the Ehrenfest theorem can not be directly
applicable.

Recently, possible schemes of the experimental setup to explore the
quantum dynamics of a mesoscopic ring threaded by a time-dependent
magnetic flux have been proposed by either capacitively coupling the
ring to a electronic reservoir~\cite{Arrachea02} or applying two
shaped time-delayed pulses~\cite{Matos05}.  The quantum dynamics in
a time-dependent flux-driven ring should be achievable within recent
fabrication capability.

\sect{Acknowledgment}
The authors are grateful to D.~H. Lin, Y.~M. Kao, and M. Moskalets
for discussion of the results. This work was supported by the
National Science Council, the National Central University, and the
National Center for Theoretical Sciences in Taiwan.

\end{document}